# Simulating Performance of a BitTorrent-based P2P TV System

Arkadiusz Biernacki

Silesian University of Technology
Institute of Computer Science, Gliwice, Poland
arkadiusz.biernacki@polsl.pl

**Abstract.** In this paper we describe a prototype of a simulation framework and some ideas which are to be used to study performance of a P2P TV system in a controllable and adjustable environment. We created a simplified model describing live video distribution in a P2P TV system. Using the model we analyse how some of the system parameters influence its behaviour. We present the preliminary results obtained at different granularity levels of measurements, describing the macroscopic system performance as well as the performance of its individual components.

## 1 Introduction

Traditional Internet TV services based on a simple unicast approach are restricted to moderate numbers of clients. The overwhelming resource requirements make these solution impossible when the number of users grows to millions. By multiplying servers and creating a content distribution network (CDN), the solution will scale only to a larger audience with regards to the number of deployed servers which may be limited by infrastructure costs. Finally, the lack of widespread deployment of IP-multicast limits the availability and scope of this solution for a TV service on the Internet scale. Therefore the use of P2P overlay networks to deliver live television in the Internet (P2P TV) is achieving popularity and has been considered as a promising alternative to IP unicast and multicast models [1]. The raising popularity of this solution is confirmed by the amount of new P2P TV applications that have became available, amongst them: PPLive, SOPCast, Tvants, TVUPlayer, Joost, Babelgum, Zattoo, and by constantly increasing amount of their users. As the P2P TV is not without drawbacks, currently the popularity are gaining solutions combining multicast, CDN and P2P approaches. However in certain performance evaluation scenarios, the components of such hybrid systems can be considered separately.

The nodes in a P2P TV network, called peers, self-organize themselves to act both as clients and servers to exchange TV content between themselves. As a result, with increasing number of network peers the number of servers in a network also increases leading to a smoother exchange process. Consequently, this approach has the potential to scale with a group size, as greater demand also generates more resources. Thus it seems important to have an insight into some of performance aspect of such system. Since the most widely deployed commercial P2P TV software mentioned above have closed architecture and are proprietary, only an experimental behavioural (black-box) characterisation of such systems is in general possible. Reverse engineering of these systems may be costly and not give answers to all nurturing questions regarding their behaviour.

To avoid these disadvantages, we prepared an OMNeT++ based P2P TV simulation by means of which we observed how macroscopic behaviour of the system is influenced by its internal structure and rules describing the interactions between its elements. Our aim is to create a simulation framework for rapid P2P TV systems prototyping which will serve as a common platform for running and comparing different solutions. At the current stage of our works, the simulation will enable us to have a rough insight into some issues emerging while prototyping P2P TV systems and applications, e.g. what influence on the system have: performance and capacity of particular peers, audio-video stream forwarding capabilities of the peers, number of sources which are emitting the content and some properties of an overlay network topology.

## 2 Previous works

While P2P TV has drawn interest from researchers, most studies have concentrated on measurements of real world data traces and their statistical analysis [2][3][4], reverse engineering [5][6][7], performance comparison of different systems [8][9][10][11], or crawling P2P systems [12]. These approaches has significant advantages with respect to the reliability of the extracted results however collecting representative global information from the complex and dynamic P2P overlay network is not simple and the data gathered may be incomplete. Thus some research works tend towards examination of the P2P system in controllable simulation environment. One of the obstacle in the way is availability of proper tools for this kind of experiments. Whereas there are simulators or simulation libraries dedicated for P2P systems, they rarely directly support simulation of P2P TV solutions. Nonetheless some of these tools may be adopted for this purpose. P2PTVSim is the P2P TV simulator initially developed by Polytechnic University of Turin, but then evolved to a more general P2P simulator and is used by P2P TV researchers [13]. The tool mainly simulates the flow of data traffic through a network of interconnected peers and aims to evaluate several, mainly push-based, chunk scheduling algorithms used for video streaming. It uses mostly chunk diffusion delay and an amount of lost chunks as an algorithm efficiency criterion. The simulator implements simplified coarse-grained representations of the underlying network and several predefined overlay topologies. Except of the type of simulated scheduling algorithm, one can configure such parameters as peers number, number of random neighbours that each peer is able to connect to, set of different bandwidth-classes of peers and upload bandwidth of the source peer. As a result, a user obtains information amongst others about the delay of every chunk and a statistics of uploaded and downloaded chunks for every peer. Another related simulator, SSSim [14], is dedicated for comparison of the streaming performance of different chunk and peer scheduling algorithms. The simulator is based on some simplifying assumptions, amongst them: all peers are synchronised and have the same output bandwidth and infinite input bandwidth. Additionally peers have the possibility to know the internal state of all the other peers in the system. Therefore both P2PTV and SSSim are primarily dedicated for fast prototyping of streaming algorithms for P2P systems. PeerSim [15] is a Java P2P simulator which main purpose and goal is an exploration of messaging protocols, gossiping and epidemic diffusion, and has not been dedicated to handle a continuous stream of information, so that it presents some scalability problems especially in streaming dimension (number of chunks), which characterise P2P TV applications. However, due to its focus on messaging and the large number of already implemented gossiping algorithms, it can be used to explore details of signalling traffic and overlay management of P2P TV systems. Some analysis of P2P TV systems were performed using PlanetLab environment, amongst them is [16], where the authors monitored SopCast application placed in 70 PlanetLab nodes.

The above mentioned simulators are flow based or application level simulators which means they work at an application level and disregard parts of underlay network stack. This give them good time and memory efficiency but simultaneously providing drawbacks in terms of simulation reality because of many simplified assumptions introduced e.g. knowledge of a system global state by its peers, coarse-grained representation of lower network layers or not including them at all, lack of the control traffic implementation. All the simulators are standalone, they are not based on any other general discrete simulator, thus simulating hybrid systems providing live video and TV may somehow be difficult.

Hence the main contribution of this work is presentation of our simulation solution based on OMNeT++, which in an assumption should take into account more realistic network scenarios. We plan to provide support for underlay and overlay layers using for this purpose INET and OverSim [17] libraries. Thus our framework will focus rather on more detailed implementation of a single TV streaming protocol than coarse-grained comparison of several protocols as in the works cited above. The work is an extension to [18].

## 3 Simulator

Due to number of different approaches to P2P TV distribution, it is impossible to simulate them all. Thus in our research we focus on the most popular and verified in practise BitTorrent-based (BT-based) solutions. Originally BT has shown to be very efficient in distributing large files, however currently it is also used for distributing video streams [19]. Amongst others, the most popular P2P TV systems like mentioned PPLive or SopCast are based on the BT protocol, with a channel selection bootstrap from a Web tracker, and peer download/upload video in chunks from/to multiple peers [12].

Generally there are four different types of components in a typical BT-based P2P TV system (e.g. GoalBit [20]), see Fig. 1(a) – a source (broadcaster), peers, superpeers, and a tracker. The source is responsible for content to be distributed i.e. for getting it from a storage or TV camera and to put it into the platform. The superpeers are highly available peers with a large capacity helping in the distribution of the content. The peers, representing the large majority of nodes in the system, are final users, who connect themselves to the streams for playout. The tracker is in charge of management of the peers in the system. For each TV channel the tracker stores a reference to the peers connected to it. The peer periodically learns about other peers connecting the tracker and parsing a peer list returned. The peer joins the swarm by establishing connections with some others peers.

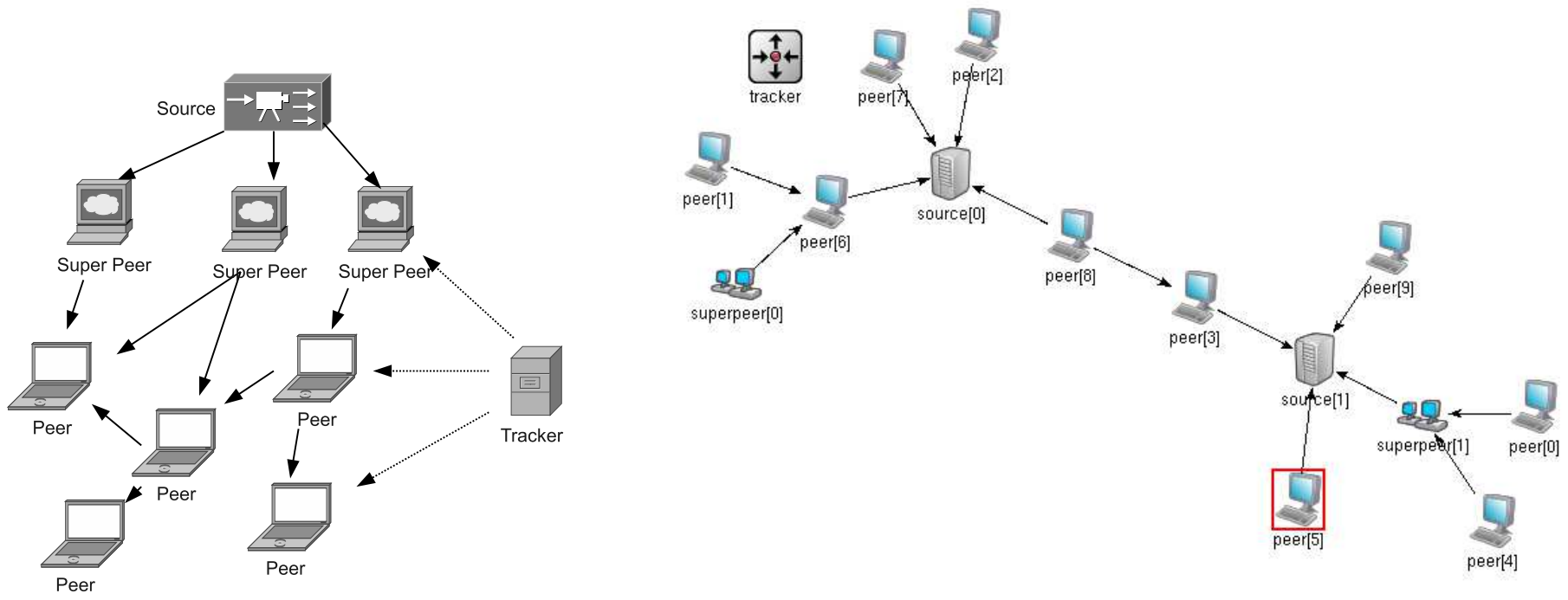

(a) BT-based P2P TV architecture, source [20]

(b) Simulation

**Fig. 1.** System components

All the above components were implemented in our simulator, however with functionality reduced to establishing and tearing down connections, monitoring the bandwidth status and basic control traffic exchange, see Figure 1(b). While modelling data traffic we do not focus on individual packets (chunks), which does not make much sense in case of P2P content networks, but we model a stream of chunks. Every network node is described by several attributes, amongst them: performance, forwarding ability of audio-video stream, a maximum number of incoming and outgoing connections. Other globally controllable parameters involve a number of transmitted TV channels, a number of network nodes: peers, superpeers and sources. So far we have ignored an influence of underlying and overlying protocols and focused on an application logic. Another simplification is that simulation currently supports only a single tracker and does not use trackless DHT mode. All messages are responded immediately which implies that their simulation processing time is zero. In our approach we used an event driven approach, where a scheduler maintains a list of simulation events.

## 4 Theoretical model and the experiment assumptions

For most of our experiment we simulated the system composed of a 1200 peers nodes, variable number of superpeers and sources, and a single TV channel. All the implemented overlay connections between the nodes in our simulation were directional. By outgoing connection for node $P_a$ we denote a link between node $P_a$ and any other node $P_i$ by which $P_a$ downloads data from $P_i$. By an incoming connection for node $P_a$ we denote a link between node $P_a$ and any other node $P_i$ by which $P_a$ uploads the data to $P_i$. When two nodes are connected by either type of the link we call them neighbours. We measure performance of single system nodes in terms of downloading and uploading content goodput (the application level throughput) which can be interpreted as packet stream intensity that a node is able to download or upload in a certain time unit.

The downloading goodput $P_a^{DG}$ of node $P_a$ is a sum of uploading goodput $P_i^{UG}$ of all the directly connected nodes $P_i$ to node $P_a$ via its outgoing connections limited by its maximum download goodput $P_a^{DG_{max}}$:

$$P_a^{DG} = \min(\sum_{i=1}^{n} P_i^{UG}, P_a^{DG_{max}}). \tag{1}$$

For the source nodes $P_a^{DG} = P_a^{DG_{max}}$ because they do not download any data from other nodes in the network.

The upload goodput for peers and superpeers is defined as

$$P_a^{UG} = RP_a^{DG}/n \quad 1 \leq n \leq N, \tag{2}$$

where $R$ is the audio-video stream repeatability coefficient and $n$ is number of incoming connections, see also Fig. 2. The download goodput of a single peer depends on $P_a^{DG_{max}}$ parameter which in practise may be related to underlay network performance in which this peer is embedded. Our assumption is that a user is able to watch the channel if $P_a^{DG}$ of its peer is greater than 0.5, otherwise he disconnects for certain time specified in the simulation configuration. Consequently, in our simulation we set $P_a^{DG_{max}}$ parameter to a random value generated uniformly from a range 0.5 to 1.0. Upload goodput depends on an ability for repeating the received stream and may be interpreted as a result of asymmetry in download and upload capabilities of the underlying network. Generally, we assumed that for the common peers $R \leq 1$ and $P_a^{DG_{max}} = 1$, for the superpeer $R \geq 1$ and $P_a^{DG_{max}} = 2$, and for the source peers $R = 1$ and $P_a^{DG_{max}} = 1.5$. Hence the superpeers are treated as servers with good upload abilities which can simultaneously distribute the same content to many peers using unicast method. Our implementation provides configurable upper bounds for the number of established both outgoing and incoming connections, which is regulated by the $N$ parameter for every network node.

To clarify the theory behind the above mentioned coefficients we present a simple example. Our network has a topology as presented on Fig. 2. As all the nodes are common peers so, as we assumed earlier, for all of them $P_i^{DG_{max}}$ parameter has a random value generated uniformly from a range 0.5 to 1.0. For the purpose of this example and simplicity of computation let us assume that for all five nodes $P_i^{DG_{max}} = 0.8, i = 1 \ldots 5$ and the repeatability coefficient $R = 1$. Let us also assume that upload goodput of the nodes 1 and 2 are $P_1^{UG} = 0.4$ and $P_2^{UG} = 0.3$ respectively. Thus, according to (1), the download goodput of node 3 will be

$$P_3^{DG} = \min(P_1^{UG} + P_2^{UG}, P_3^{DG_{max}}) = \min(0.4 + 0.3, 0.8) = 0.7.$$

Node 3 forwards the stream to nodes 4 and 5 (so in this case $n = 2$) and, according to (2), its upload goodput is

$$P_3^{UG} = RP_3^{DG}/n = 0.35.$$

Hence the nodes 4 and 5 receive from node 3 stream with upload coefficient 0.35.

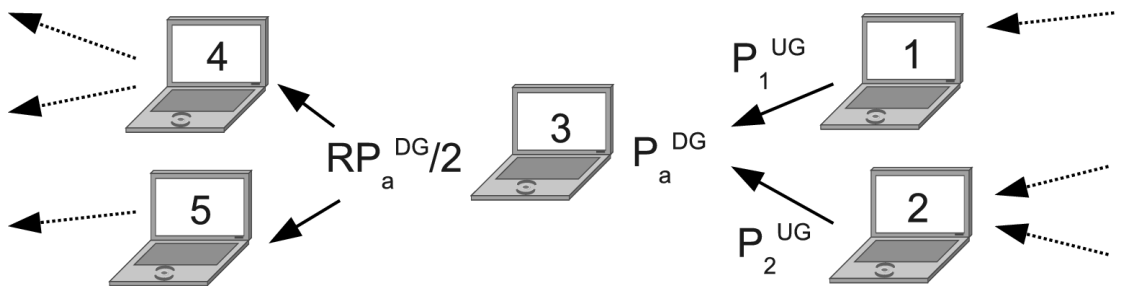

**Fig. 2.** Computation of the download and upload goodput

The system topology is created dynamically from scratch. We implemented a random churn generator. In fixed time intervals a random number is drawn and depending on this number, a random node is either added or deleted. Peers and superpeers periodically query a tracker to obtain lists of their neighbours and the neighbours' parameters. Every peer creates a ranking of neighbouring peers and tries to connect to a peer with the highest upload goodput (2). After the connection, the peer monitors its download goodput in certain time intervals which is a parameter of the simulation. If the number of connections is equal to the maximum allowed number of connections, the peer disconnects from a neighbour which has the worst upload goodput and tries to reconnect to a better one, thus making network topology constantly evolve. At the current stage we do not implement the chocking – a popular BitTorrent mechanism involving temporary refusal to upload.

The main purpose of our work was to examine how the above mentioned parameters: the number of superpeers, the number of sources, $R$, and $N$ will influence on its goodput represented by $P_a^{DG}$ and $P_a^{UG}$. The results were presented at three different levels: the analysis of global values, where we summarised and aggregated multiple peers behaviour in a function of a few system parameters;

the analysis of average values, where we compared the behaviour of a several peers; and the transient behaviour of single peers, where we monitored the behaviour of selected peers in the function of time. These simple analyses could be helpful for the P2P TV designers and developers facing the question of how many certain system special components like sources or superpeers should be used to provide the system users acceptable quality of audio-video transmission. From the other side, it can be also useful for the system access control – having defined the system infrastructure the system designer can assess the amount of users who can access the system simultaneously without degradation of its performance below an acceptable level.

## 5 Results

In the first experiments we obtained a macroscopic view of the system studying how upload to download ratio $R$ (2) affects the system performance. Our system had 4 sources, 16 superpeers, the number of allowed connection for peers was set to 8 and, as it was mentioned earlier, the system had 1200 peers. We concluded that the decrease of $R$ parameter from 1 to 0.9 provided relatively low impact on system performance, see Fig. 3(a), nonetheless another decrees from 0.9 to 0.8 resulted in dramatic reduction of our system performance measured both as upload and download goodput. An interesting question arises: for which value range of $R$ the performance of the system deteriorates the fastest?

In the second analysis we examine how the number of maximum allowed connections $N$ (2) influence the system performance using the same set of parameters as in the first experiment. We claim that the parameter had minor impact on system download goodput however it affected upload goodput, see Fig. 3(b). Such behaviour has simple explanation, according to (2) there is an inverse proportion between the upload goodput and the number of incoming connections, thus with the increasing number of connections, the upload goodput decreases.

Increasing number of sources from 2 to 8 gradually increased the system goodput, see Fig. 3(c). However instead of adding content sources, which may be problematic due to synchronization of content transmission, we can increase the system performance by adding more superpeers. The system efficiency is quite sensitive to a number of superpeers, however adding more than 16 superpeers did not lead to any further improvement in the system goodput, see Fig. 3(d).

It should be recalled that the aforementioned values strictly depend on the others of the experiment parameters, amongst them the amount of peers.

The results of the simulation can be further analysed on more details levels. We extended the results presented on Fig. 3(a) obtaining the download goodput for a several selected peers separately, see Fig. 4(a). The values varied amongst the peers suggesting that in certain cases the global analysis might not be enough to achieve a reliable view of the examined system. In accordance with Fig. 3(a) there was a noticeable difference when stream repeatability coefficient $R$ drops from 0.9 to 0.8. We could also observe that the last mentioned value of the

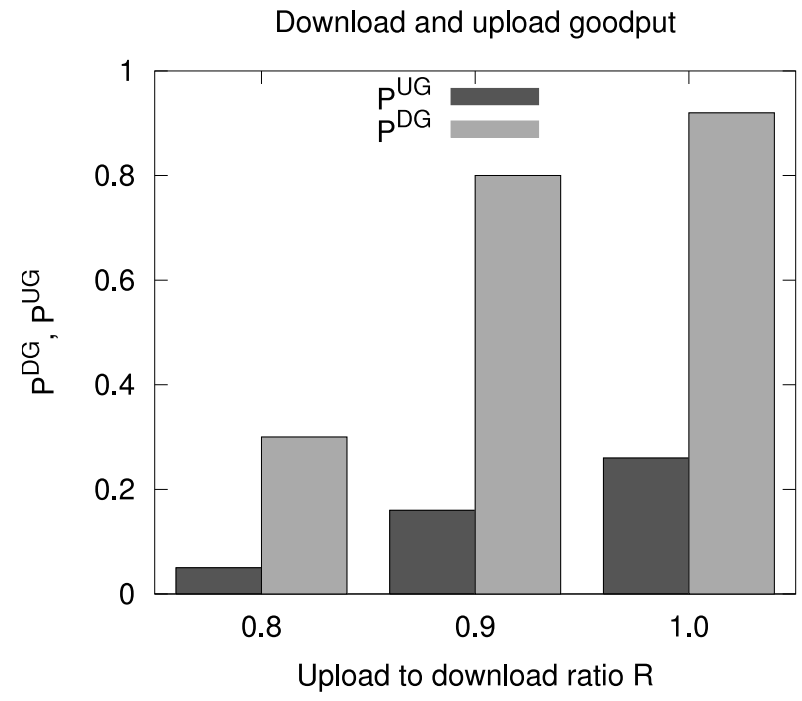

(a) Upload to download ratio $R$, $N$=8, 4 sources, 16 superpeers

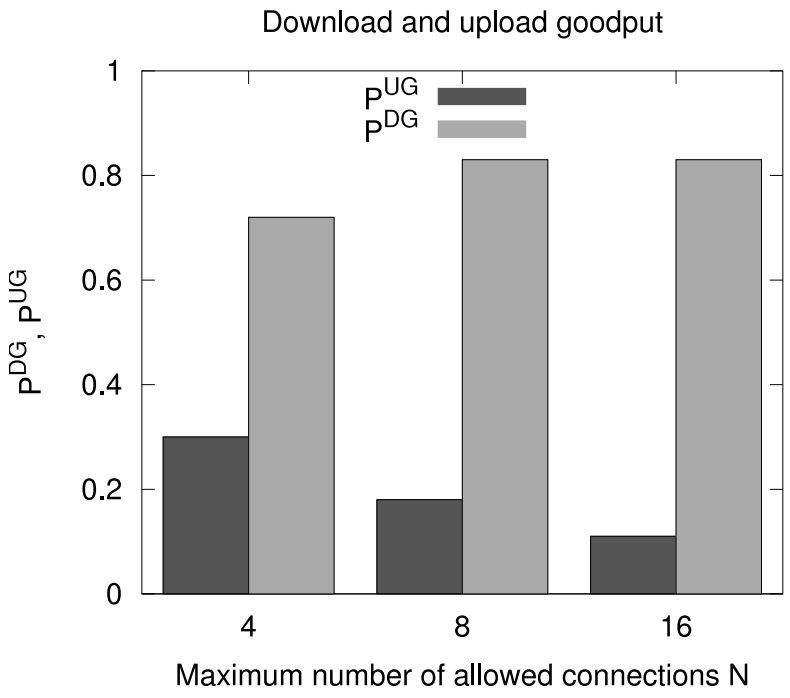

(b) Number of allowed connections, $R$=0.9, 4 sources, 16 superpeers

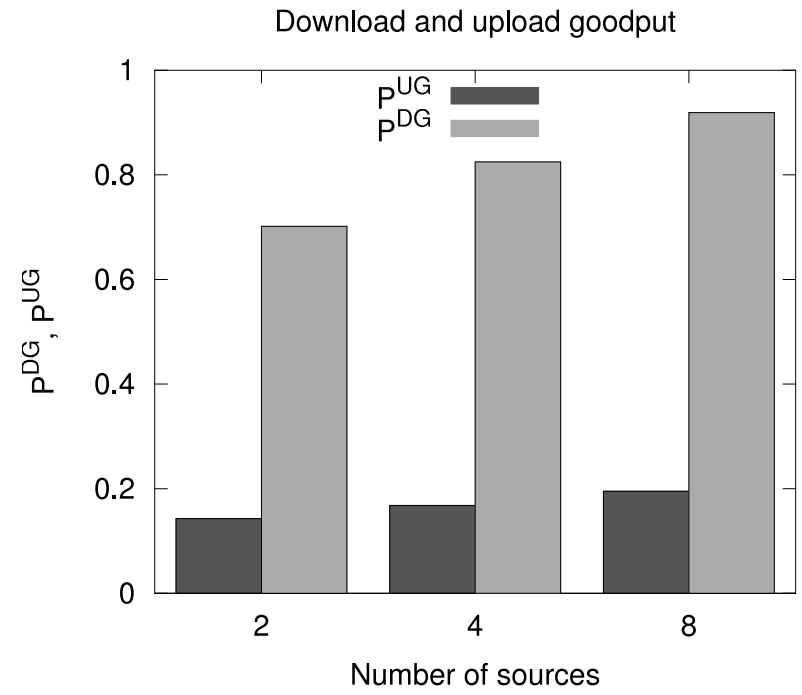

(c) Number of sources, $N$=8, $R$=0.9, 16 superpeers

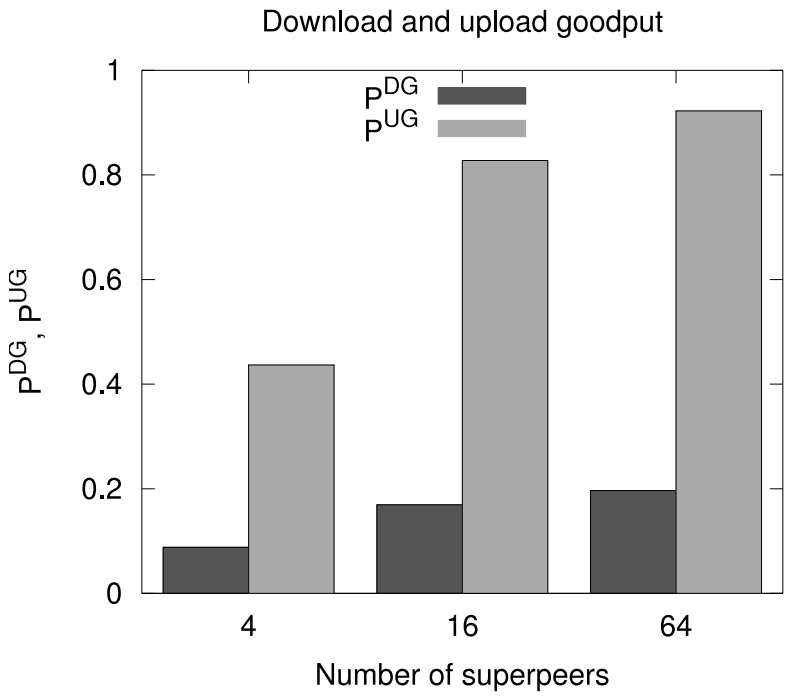

(d) Number of superpeers, $N$=8, $R$=0.9, 4 sources

**Fig. 3.** Aggregate goodput measurements in a function of different simulation parameters

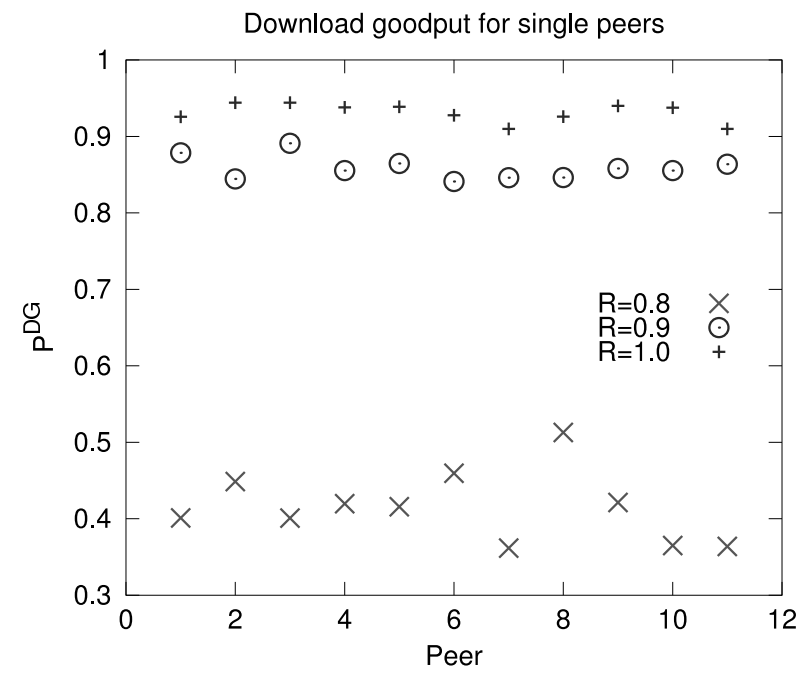

(a) Multiple peers comparison, $N$=6, 4 sources, 16 superpeers

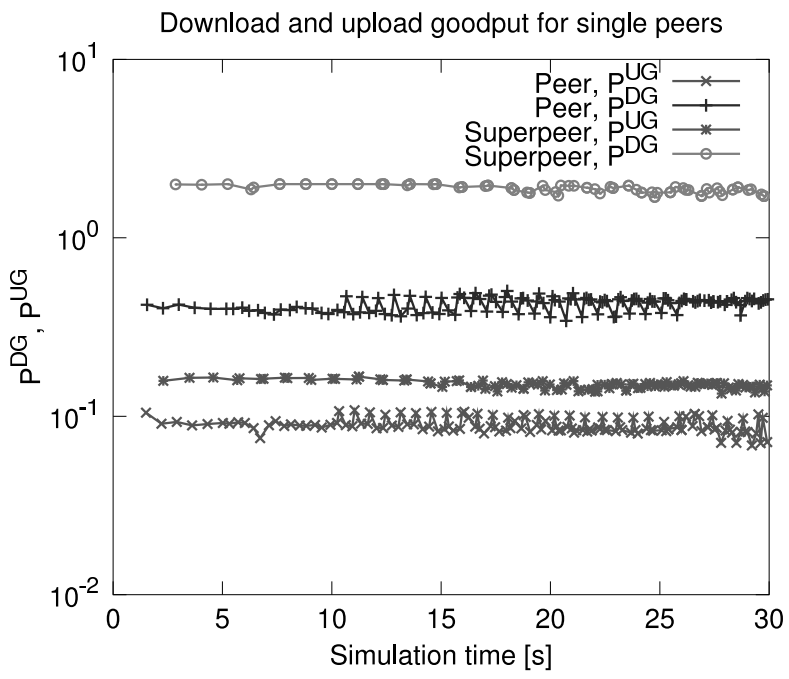

(b) Transient analysis, $N$=6, $R$=0.9, 4 sources, 16 superpeers

**Fig. 4.** System goodput – detailed level

repeatability coefficient tiggered the highest variation of download goodput for the examined peers. Taking one step further we were able to analysis a single scenario from Fig. 4(a) in terms of a peer transient behaviour. On Fig. 4(b) we presented download (1) and upload (2) goodput for randomly selected peer and superpeer. As expected the download goodput of the superpeer clearly surpasses the goodput of the common peer. However, the difference between upload goodput of the superpeer and the peer was not as dramatic, which proved that the the superpeer fulfilled its role distributing its content. Both upload and download goodput were characterized by small oscillations even though we did not observe any huge fluctuations.

## 6 Conclusions

In this paper we have presented a draft of an implementation of the BT-based P2P TV system using the OMNeT++ simulation environment. The preliminary results presented in this paper indicate that using the proposed simulator we are able to perform small-scale simulations of the simplified system showing its behaviour in micro- and macroscopic scale, which potentially may be helpful for fast prototyping of this kind of P2P TV systems. The results were obtained assuming a number of simplifications in our modelled system, especially concerning chunk selection, chunk buffer management and underlay network modelling. However our approach, i.e. a choice of popular discrete event simulator for the implementation, allows us to gradually incorporate further details. In the future we plan to model the P2P TV exchange protocol with greater attention to its details and provide support for underlying and overlay layers using INET and OverSim libraries. In an assumption our simulation framework should not only be dedicated to P2P TV but also should provide the researches possibilities of testing hybrid scenarios involving multicast and CDN solutions.